\documentclass[aps,prb,twocolumn,showpacs,amsmath,amssymb]{revtex4-1}
\usepackage{graphicx}% I deleted here the epsf package
\usepackage{bm}
\usepackage{amsmath}

\newcommand{\be}{\begin{equation}}
\newcommand{\ee}{\end{equation}}
\newcommand{\bea}{\begin{eqnarray}}
\newcommand{\eea}{\end{eqnarray}}
\newcommand{\up}{\uparrow}
\newcommand{\down}{\downarrow}
\newcommand{\bwt}{\begin{widetext}}
\newcommand{\ewt}{\end{widetext}}
\newcommand{\ham}{\mathcal{H}}

\newcommand{\ra}{\rangle}
\newcommand{\la}{\langle}
\newcommand{\bsb}{\begin{subarray}}
\newcommand{\esb}{\end{subarray}}
\newcommand{\largem}{\!\!}

\newcommand{\eins}{\mbox{$1 \hspace{-1.0mm} {\bf l}$}}

\newcommand{\vecv}[2]{
\left(\largem
 \begin{tabular}{c}
  $#1$ \\
  $#2$
  \end{tabular}
  \largem
\right)
}

\newcommand{\vech}[2]{
\left(\largem
 \begin{tabular}{c}
  $#1$ \! $#2$
  \end{tabular}
  \largem
\right)
}

\newcommand{\mat}[4]{
\left(
\largem
 \begin{tabular}{cc}
  $#1$ & $#2$ \\
  $#3$ & $#4$
  \end{tabular}
  \largem
\right)
}

\begin{document}

\title{Tight-binding description of intrinsic superconducting correlations in multilayer graphene}

\author{W.A. Mu\~noz}
\email{WilliamArmando.Munoz@ua.ac.be}
\author{L. Covaci}
\email{lucian@covaci.org}
\author{F.M. Peeters}
\email{Francois.Peeters@ua.ac.be}
\affiliation{Departement Fysica, Universiteit Antwerpen, Groenenborgerlaan 171, B-2020 Antwerpen, Belgium}

\date{\today}

\begin{abstract}
Using highly efficient GPU-based simulations of the tight-binding Bogoliubov-de Gennes equations we solve
self-consistently for the  pair correlation in rhombohedral (ABC) and Bernal (ABA) multilayer
graphene by considering a finite intrinsic s-wave pairing potential. We find that the two
different stacking configurations have opposite bulk/surface behavior for the order parameter. Surface superconductivity is robust for ABC stacked multilayer graphene even at very low
pairing potentials for which the bulk order parameter vanishes, in agreement with a recent analytical approach. In contrast, for Bernal stacked multilayer graphene, we find that the order
parameter is always suppressed at the surface and that there exists a critical value for the
pairing potential below which no superconducting order is achieved. We considered different doping
scenarios and find that homogeneous doping strongly suppresses surface
superconductivity while non-homogeneous field-induced doping has a much weaker effect on the
superconducting order parameter. For multilayer structures with hybrid stacking (ABC and ABA) we find that when the thickness of each region is small (few layers),
high-temperature surface superconductivity survives throughout the bulk due to the proximity effect
between ABC/ABA interfaces where the order parameter is enhanced.
\end{abstract}


% 73.43.-f: QHE
% 73.23.-b: Electronic transport in mesoscopic systems
% 73.63.-b: Electronic transport in nanoscale materials and structures

\maketitle

%%%%%%%%%%%%%%%%%%%%%%%%%%%%%%%%%%% Introduction %%%%%%%%%%%%%%%%%%%%%%%%%%%%%%%

\section{Introduction}
Superconducting correlations in graphene-based structures have been the focus of intensive theoretical and experimental research
even before  graphene became one of the most important topics in condensed matter physics.
Following last decade experimental evidences reporting hints of superconductivity       
behavior in graphite \cite{kopelevich_ferromagnetic-_2000} and graphite intercalated compounds 
\cite{silva_indication_2001, kopelevich_high-temperature_2004, 
moehlecke_interaction_2004, 
lamura_experimental_2006},
a considerable amount of theoretical studies have been devoted to provide a clear understanding about
possible mechanisms that could induce intrinsic superconducting states in single and multilayer graphene
\cite{gonzalez_electron-electron_2001, 
uchoa_superconducting_2007,
black-schaffer_resonating_2007,
roy_unconventional_2010, 
pathak_possible_2010, 
kopnin_high-temperature_2011,
profeta_phonon-mediated_2012,
nandkishore_chiral_2012}. 
More recently experimental investigations have  reported intriguing traces of high-temperature superconducting
behavior in highly oriented pyrolytic graphite (HOPG) samples
\cite{
larkins_indications_2011,
scheike_can_2012, ballestar_josephson-coupled_2013}, feeding speculations about the existence of intrinsic superconducting
correlations in graphite and graphite-based compounds.
Despite the fact that most of these experimental evidence suggests  that superconductivity in graphite compounds appears due
to external causes, several theoretical studies reveal the possibility of inducing  superconductivity
in graphite by considering unconventional symmetry of the order parameter
\cite{gonzalez_electron-electron_2001,
black-schaffer_resonating_2007,
uchoa_superconducting_2007,
roy_unconventional_2010,
nandkishore_chiral_2012}. 
However, these calculations show that superconductivity becomes stable after considering disorder 
\cite{gonzalez_electron-electron_2001} or high doping in the graphene
layers \cite{uchoa_superconducting_2007,black-schaffer_resonating_2007} while surface superconductivity appears to be stable in
clean rhombohedral graphite in the absence of external doping \cite{kopnin_high-temperature_2011}.
Considering that most of these calculations are based on a reduced Hamiltonian or are performed within two-dimensional models, by
ignoring the interplanar hopping, a numerical description of the superconducting correlation in multilayer graphene is urgently
needed.

In view of this, we provide in the following a numerical description of intrinsic superconductivity in multilayer graphene at 
the tight-binding level.
Following Ref. \onlinecite{kopnin_high-temperature_2011}, we consider a simple s-wave pairing  symmetry in a ABC (or rhombohedral)
stacking multilayered graphene structure.
Calculations are also performed for ABA (or Bernal) stacking where its quadratic low-energy  band
structure \cite{partoens_graphene_2006} shows a remarkable difference from the  $|p|^N$ momentum dependent band structure seen in ABC
multilayer \cite{guinea_electronic_2006} structures. In particular we are interested in the limit of large number of layers ($N$) where
the lower-energy band in the rhombohedral case is flat over a large region in k-space signaling the suppression of the kinetic 
energy and therefore resulting in  strong effect from interactions.
In addition, because of the sensitive stacking dependent band structure in multilayer graphene, we
also considered hybrid stacking cases. It is known that exfoliated few-layer graphite samples are usually found
to exhibit very stable Bernal stacking but often also display rhombohedral structures in part of the sample
\cite{lui_imaging_2011,ping_layer_2012}. It is worth mentioning that during the preparation of the manuscript, new experimental results revealed the existence of superconducting correlations at two-dimensional interfaces that appear when angle misalignments about the
\textit{c}-axis exist in HOPG \cite{ballestar_josephson-coupled_2013}.

By using highly efficient graphics card (GPU) simulations of the tight-binding Bogoliubov-de-Gennes equations, we are able to solve self-consistently
for the pair
correlation in  multilayer graphene by considering both planar and inter-planar coupling between nearest neighbors.
Translational
invariance is assumed along the 2-dimensional direction within the graphene sheets. In this way, an adequate description for the profile of
superconducting correlations along the direction perpendicular to the graphene layers is achieved. 

Our results confirm the main features of recent analytical approaches for ABC rhombohedral graphite where an enhancement of
surface superconductivity, with  respect to  its bulk analog,  was recently predicted \cite{kopnin_high-temperature_2011}.
 We find that the opposite behavior is present in ABA stacking where bulk dominates over surface superconductivity. This fact requires that a strong 
pairing potential is needed in order to obtain a non-zero pair correlation for the ABA case. In contrast, in the ABC case, a lower pairing potential
is sufficient to induce a large pair amplitude at the surface. In addition, we show how doping influences surface superconductivity, i.e. it
strongly suppresses it.

The paper is organized as follows. In Sec.~\ref{model} we introduce the model and the numerical approach that we use. In Sec.~\ref{results} we
present and discuss the results of our numerical calculations. Finally, we briefly summarize our findings in Sec.~\ref{conclusions}.
\begin{figure}[bbb]
%\vspace{-0.2cm}
\includegraphics[width=\columnwidth]{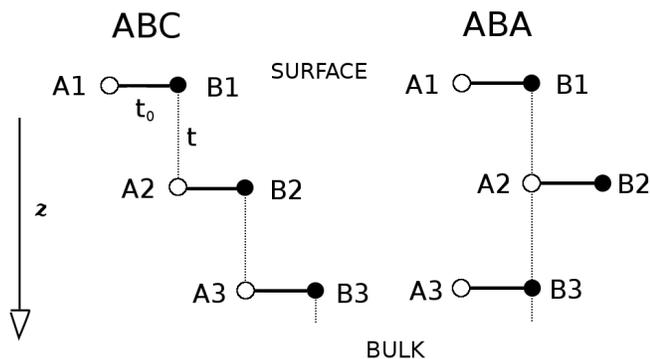}
%\vspace{-0.2cm}
\caption{\label{layout} Schematic layout of the side view of a multilayer graphene structure stacked in two different
configurations: Rhombohedral or ABC (left) and Bernal or ABA (right). Sublattices A and B are coupled by $t_0$ within the
same layer, while interlayer coupling is described by $t$. Integer coordinate $z$ correspond to the index layer. 
 }
\end{figure}
\section{Model and calculation approach}
\label{model}
Superconducting correlations in multilayered graphene structures are described by the following mean-field single particle
Hamiltonian written in Nambu space:
\begin{eqnarray}
\ham=\sum_{ \substack{<l,m> \\ <i,j>} } \vech{c_{l\up}^{i\dagger}}{c_{l\down}^{i}}
\mat{\hat{\ham}_{lm}^{ij}}{\Delta^i_l\delta_{ij}\delta_{lm}}{\Delta_l^{i\ast}\delta_{ij}\delta_{lm}}{-\hat{\ham}_{lm}^{ij\dagger}
}
 \vecv{c_{m\up}^{j}}{c_{m\down}^{j\dagger}} 
\label{ham}
\end{eqnarray}
where the summation, $\langle i,j \rangle$, is done over nearest neighbors within each layer while the summation, $\langle l,m \rangle$, is done for  adjacent layers. The
non-diagonal elements $\Delta_l^i$ correspond to the s-wave superconducting order parameter at the atomic site $i$ in
 layer $l$ while the diagonal elements $\ham^{ij}_{lm}$ are the normal state components of the Hamiltonian:
\begin{eqnarray}
\label{ham0}
\hat{\ham}^{ij}_{lm}=[-t_0(1-\delta_{i,j})-\mu_l\delta_{i,j}]\delta_{l,m}-t(\delta_{l,m+1}+\delta_{l,m-1}) \;\;\;
\end{eqnarray}
where $\mu_l$ is the chemical potential and, according to the layout of Fig. \ref{layout}, nearest-neighbors sublattices A
and B are coupled within the layers by the hopping parameter
$t_0\approx$2.8eV while $t=0.1t_0$ describes the hopping  which couples A sites with the
 nearest B sites in the adjacent (upper and lower) layers.
Bernal and rhombohedral stacking are defined according to the vertical symmetry along the z-axis as shown in
Fig. \ref{layout}.
Due to this symmetry, rhombohedral stacking allows us to reduce the description of the superconducting
parameter to one of the sublattices, whereas the other sublattice can be deduced from a mirror reflection transformation.
For Bernal stacking, we consider a more practical way of  sorting sites inside the supercell as dimmer sites, which
correspond to the sublattices coupled by the interlayer hopping $t$, and no-dimmer sites or sublattices which does not participate
in the coupling between adjacent graphene sheets.
As we  previously pointed out in a recent work \cite{munoz_tight-binding_2012}, the pair correlation behaves differently in
these inequivalent sites because at dimmer sites the density of states vanishes around the Dirac point while
being finite at no-dimmer sites where the formation of Andreev states is more feasible.

We will not specify the origin of superconductivity in the multilayer graphene structure, but rather assume $\Delta_l^i=U\la c_{l\up}^i c_{l\down}^i \ra$
to be a conventional s-wave symmetric order parameter and the pairing potential $U$  is fixed and homogeneous in
the whole structure.
Under this assumption and considering translational invariance along the transversal directions, we solve self-consistently
for the amplitude of the  pair correlation
 $|\la c_{l\up}^i c_{l\down}^i \ra|$ for the sublattice A (or both in the case of N-Bernal stacked layers for N
odd),  in the $z$ direction. We have considered graphene multi-layer supercells of size 42nm$\times$25nm$\times N$ such that the order parameter is converged and no additional momentum summations in the parallel direction are needed.
Instead of a direct diagonalization of the Hamiltonian we performed the self-consistent
mean-field calculation through a numerical approximation of the Gorkov Green's function by using the
Chebyshev-Bogoliubov-de-Gennes method \cite{covaci_efficient_2010, covaci_superconducting_2011}. Both, the normal
and anomalous Gorkov Green's function, can be
approximated by a
superposition of a finite number of Chebyshev polynomials as follows:
\begin{eqnarray}
\label{greenseries}
\bar{G}_{ijlm}^{1\alpha}(\tilde{\omega})=\frac{-2i}{\sqrt{1-\tilde{\omega}^2}}\left[\sum_{n=0}^N
a_{ijlm}^{1\alpha}(n)e^{-in\arccos(\tilde{\omega})}\right],
\end{eqnarray}
where the expansion coefficients for the diagonal, or normal ($\alpha=1$), and
the off-diagonal, or anomalous ($\alpha=2$), components of
the 2$\times$2 Green function are defined respectively as
\cite{covaci_efficient_2010}:
\begin{eqnarray}
\label{normal}
a^{11}_{ijlm}(n) = \la c_{l\up}^i\left|T_n(\ham)\right|c_{m\up}^{j\dagger}\ra \\
\label{anomalous}
a^{12}_{ijlm}(n) = \la c_{l\down}^{i\dagger}\left|T_n(\ham)\right|c_{m\up}^{j\dagger}\ra^{\ast}
\end{eqnarray}
where $T_n(x)=\arccos[n \cos(x)]$ is the Chebyshev polynomial of order $n$, which  satisfies
the following recurrence relation: $T_{n+1}(x)=2xT(x)-T_{n-1}(x)$.
Once the Hamiltonian has been normalized according to $\ham\rightarrow \tilde{\ham}=(\ham-\eins b)/a$, where the rescaling
factors are $a=(E_{max}-E_{min})/(2-\eta)$ and $b=(E_{max}+E_{min})/2$, with $\eta>0$ being a small number, the expansion coefficients
can be obtained
through the dot product   $a^{1\alpha}_{ijlm}(n) = \la \alpha|v_n\ra $, where $\la \alpha|$ are the vectors $\la 1|=\la c_{l\up}^i|$
and $\la 2|=\la c_{m\down}^{j\dagger}|$ and the recursive vector $|v_n\ra= 2\ham|v_{n-1}\ra- |v_{n-2}\ra$.
We refer the reader to  Ref. \onlinecite{covaci_efficient_2010} for more details about the numerical procedure.
Since this iterative procedure  involves mainly a successive application of the Hamiltonian matrix (\ref{ham}) on iterative vectors
$|v_n\ra$, most of the computational effort corresponds to  sparse matrix-vector multiplication which can be performed with high-efficiency by
implementing parallel computations on GPUs by using CUDA Nvidia.
We are therefore able to solve efficiently multilayered graphene structures containing typically several hundreds of thousand of atoms, which is half of the size of the
Hamiltonian matrix (\ref{ham}).
Physical quantities, like the local density of states and  the pair correlation function, can be easily
determined once the Green's functions are known: $N_l^i(\omega)=-\frac{2}{\pi}\text{Im}{G}_{iill}^{11}(\omega)$ and $\la c_{l\up}^i
c_{l\down}^i \ra = \frac{i}{2\pi}\int^{E_c}_{-E_c}{G}_{iill}^{12}(\omega)[1-2f(\omega)]d\omega$.
\section{Numerical results}
\label{results}
\begin{figure}[ttt]
%\vspace{-0.2cm}
\includegraphics[width=\columnwidth]{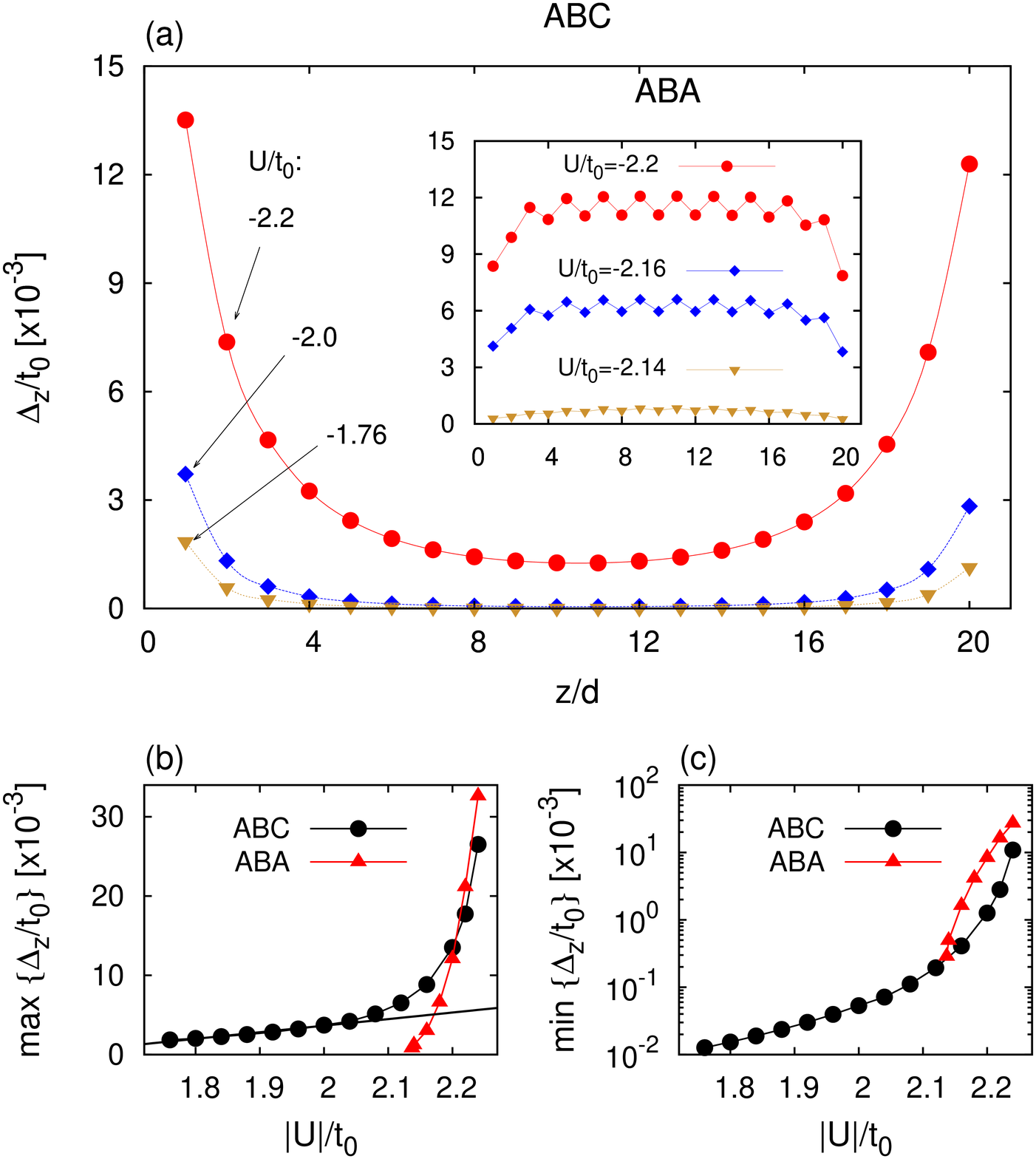}
%\vspace{-0.2cm}
\caption{\label{upro} (Color online) (a) Order parameter profile ($\Delta_z$) along the z-direction  perpendicular to the
graphene
sheets (see Fig.~\ref{layout}) for various  values of the s-wave  attractive pairing potential $U$ in ABC stacked multilayer
graphene with $N=20$ layers.
The inset shows the corresponding $\Delta_z$ profile for the ABA case where dimmer and non-dimmer sites follow different
curves for higher values of $U$. $U$-dependence of the maximum (b) and minimum (c) value of $\Delta_z$ for both ABC and ABA
stacking
configuration. $d=0.335$nm is the inter-layer distance.
}
\end{figure}
We solved self-consistently for the order parameter along the z-direction, and show in Fig.~\ref{upro}(a) the different profiles
of the order parameter ($\Delta_z$) 
for different values of the pairing potential ($U<0$) in ABC stacked graphene with $N$=20 layers. The order parameter is only
shown for the A sublattice,
while the B sublattice  value is achieved by a
mirror reflexion symmetry along z, as can be deduced from Fig.~\ref{layout}. Analog results are shown in the inset  of 
Fig.~\ref{upro}(a) for ABA stacking and different values of the pairing potential.

We notice in Fig.~\ref{upro}(a), that the superconducting order parameter at the outermost layers is larger than the
vanishing pair correlation in the bulk for all the U-values considered here. The same surface enhancement is observed when we decrease the
pairing potential such that the penetration of  the superconducting order parameter
into the bulk becomes strongly suppressed. In the limit of very low pairing potential good agreement could
be found with the analytical result previously reported in Ref.~\onlinecite{kopnin_high-temperature_2011}.

On the other hand, an opposite surface-bulk superconducting ratio is found for the Bernal
stacking configuration (see inset in Fig~\ref{upro}(a)). 
Self-consistent calculations performed for ABA show that the bulk value of the order parameter is dominant
while surface superconducting correlations are suppressed. Also, we can observed a sublattice
polarization in the $\Delta_z$ profile where pair correlation is found to be higher in non-dimmer sites as compared to dimmer
sites with
an energy difference which becomes smaller as the pairing potential is decreased. We return to this issue in a later discussion
about the density of states.

A direct comparison between the surface and the bulk value of the pair correlation is given in  Figs.~\ref{upro}(b) and
\ref{upro}(c) for both stacking configurations. 
Fig.~\ref{upro}(b) shows the maximum values of the superconducting correlation for both ABC and ABA cases for different values of $U$. According to the profiles observed in Fig.~\ref{upro}(a), the maximum value of the order parameter, max$\{\Delta_z\}$, corresponds to the surface for ABC stacking while for the ABA stacking the maximum value corresponds to the bulk non-dimmer sites.
In contrast, the log-linear representation presented in Fig.~\ref{upro}(c) shows the U-dependence of the minimum value of  the
superconducting order parameter, min$\{\Delta_z\}$, which corresponds to the bulk and surface locations for ABC and ABA,
respectively.

Two different regimes can be inferred from Figs.~\ref{upro}(b) and \ref{upro}(c), depending whether the pairing potential  U is
larger or smaller than a critical value, $|U_c|/t_0 \sim 2.14$. This value is also very close to the critical pairing for the ABA
stacking, below which superconducting correlations vanish. As $|U|$ decreases, but is still larger than $|U_c|$, the order
parameter decays exponentially for both ABC and ABA stacking, as the main contribution comes from the bulk. When $|U|<|U_c|$, for
ABC stacking, the bulk order parameter vanishes exponentially but the surface one is still finite. Even more interesting is the
fact that the surface order parameter decays only linearly and is non-zero for all the values of $|U|$ that we considered in the
simulation, down to $|U|/t_0=1.76$ giving a surface order parameter $\Delta_{max}/t_0=0.002$. The self-consistent calculation
becomes increasingly intensive as the order parameter decreases since more Chebyshev moments are needed to resolve the Green's 
function at higher and higher resolutions.

The bulk behavior resembles the superconducting critical point reported for graphene, where in the undoped case it was found  that
the critical temperature vanishes below a finite value of the s-wave pairing interaction \cite{kopnin_bcs_2008}. In contrast, the
$|U|$ dependence of the surface order parameter in the ABC stacking configuration suggests that the surface states, which form a
flat band with suppressed kinetic energy, are strongly influenced by any exponentially small interaction.\\
\begin{figure}[ttt]
%\vspace{-0.2cm}
\includegraphics[width=\columnwidth]{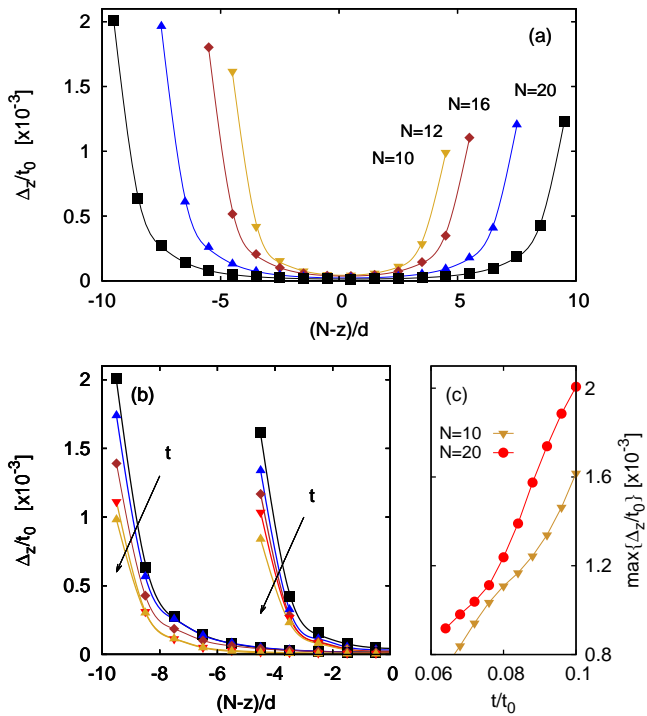}
%\vspace{-0.2cm}
\caption{ \label{npro} (Color online) (a) Order parameter profile ($\Delta_z$) for a rhombohedral multilayer graphene consisting
of various total number of layers $N$=10, 12, 16 and 20. (b) Order parameter profile for N=10 and
20 for different values of the interlayer hopping $t/t_0$=0.1, 0.092, 0.084, 0.076 and 0.068, in decreasing order as this is
indicated by
the arrows. (c) Dependence of the surface pair correlation, max$\{\Delta_z\}$, as a function of $t$. Here we used $U=-1.76t_0$.}
\end{figure}
Since the pair correlation at the surface is always enhanced for ABC stacking and survives even for lower values of the pairing
potential, we will further only also report numerical results for this stacking configuration. 

The dependence of the order parameter on the total number of graphene layers, $N$, and the interlayer coupling, $t$, is shown in
Fig.~\ref{npro} for $U$=-1.76$t_0$. 
An asymptotic enhancement of surface superconductivity is observed in Fig.~\ref{npro}(a) as the
Fermi surface size, defined by the flat band localized at the surfaces, increases with the number of layers. 
On the other hand, the decrease of the interlayer coupling leads to the suppression of the order parameter as this is seen in
Fig.~\ref{npro}(b). The evolution of the surface pair correlation as a function of $t$ is shown in Fig.~\ref{npro}(c) and
indicates an almost linear suppression as a consequence of the linear dependence of the Fermi surface size on $t$.
\cite{kopnin_high-temperature_2011}
 
\begin{figure}[ttt]
%\vspace{-0.2cm}
\includegraphics[width=\columnwidth]{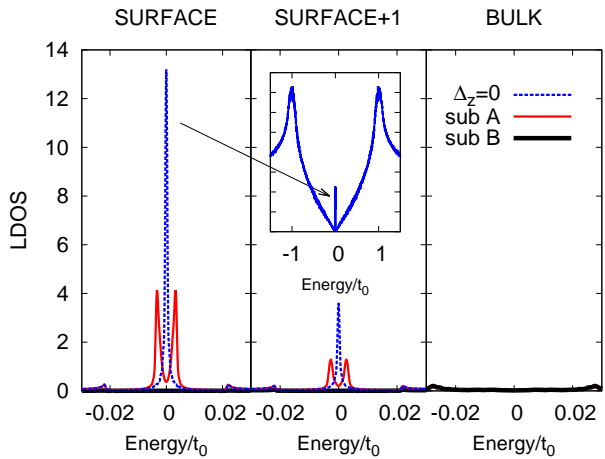}
%\vspace{-0.2cm}
\caption{ \label{ldos} (Color online) Local density of states (LDOS) showing the formation of the s-wave superconducting gap for
$U=-2.0t_0$ at both
sublattices A and B in different layers around the Fermi energy. Left and center panel show the LDOS at the surface and  its
adjacent layer while right panels show the LDOS at the bulk. The dashed line represent the corresponding
normal state LDOS which shows the localized flat band at the outermost layers. The inset of the central panel shows the surface
LDOS in ABC stacked graphene over a wider range of energies.
}
\end{figure}

In order to provide a better understanding of the peculiar behavior of the order parameter profile observed in Fig.~\ref{upro} for the rhombohedral case,  we next present the local density of states (LDOS).
Fig.~\ref{ldos} shows the LDOS in different layers for both sublattices, A and B,  within a small energy interval near the Fermi energy.
The left panel of Fig.~\ref{ldos} represents the LDOS at the surface where the superconducting order parameter is enhanced for
sublattice A. There, the normal state LDOS shows a sharp peak due to the existence of flat bands with dispersion $E\sim|p|^N$.
The corresponding wave function of these states is localized at the surface and only on sublattice A. A zoomed-out view of the LDOS, over a wider
range of energies, is shown in the inset of Fig.~\ref{ldos}.
This is very different for the B sublattice where the density of states vanishes around the Fermi energy and the 
superconducting coherence  peaks are not visible. Despite this, a non-zero solution for the order parameter is obtained for atomic
sites belonging to this sublattice as we can see  in Figs.~\ref{upro}(a) and~\ref{npro}.
This non-zero solution appears as a consequence of the proximity effect between the intra-layer neighbors A
and B sites at the surface.
We also observe less pronounced coherence peaks appearing in the LDOS of sublattice A in the layer adjacent to the surface (see
central panel of Fig.~\ref{ldos}) while
the LDOS vanishes for both sublattices in the bulk (see right panel of Fig.~\ref{ldos}).
According to this behavior of the LDOS for rhombohedral stacking, we expect that superconducting correlation will be
more stable on the surface and on few adjacent layers rather than in the bulk where the  LDOS vanishes around the Fermi energy. \\
With respect to the LDOS in the Bernal case, it is well known that a sublattice polarization
appears around the Fermi energy. A finite density of states is found for non-dimmer sites while  the density of states vanishes at
the dimmer sites  \cite{guinea_electronic_2006}.
Such a polarization allows a finite order parameter to be induced only by large s-wave pairing potentials and therefore bulk
superconductivity will not be stable for values of U  for which surface superconductivity in ABC is still finite  (see
Fig.~\ref{upro}(b)).
In this way, the suppression of  surface superconductivity seen in Fig.~\ref{upro}(c) for ABA  appears as a consequence of the
lower density of states for  surface non-dimmer sites when compared to its bulk value.
\begin{figure}[ttt]
%\vspace{-0.2cm}
\includegraphics[width=\columnwidth]{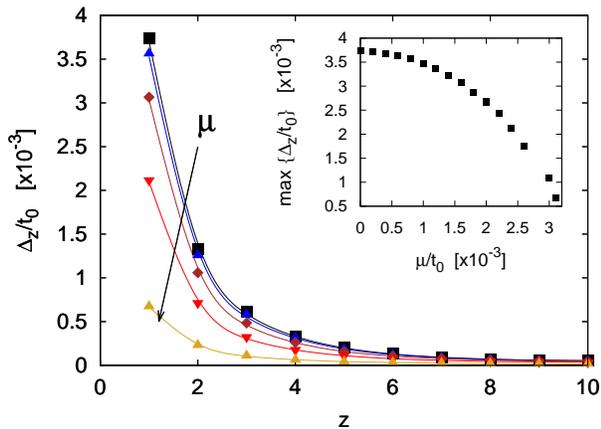}
%\vspace{-0.2cm}
\caption{ \label{dopef} (Color online) Effect of homogeneous doping on the intrinsic s-wave superconducting order parameter in a
multilayer rhombohedral graphene with $U=-2.0t_0$. We shown here half of the profile for different doping values, $\mu/t_0\times
10^{-3}$=0, 0.8,
1.6, 2.4 and 3.1, in decreasing order indicates by the arrow. Inset: Evolution of surface superconductivity (max$\{\Delta_z\}$) as
a function of doping
($\mu$) showing that an increase in $\mu$ leads to a strong suppression of the order parameter at the surface.
}
\end{figure}
\\
\begin{figure}[ttt]
%\vspace{-0.2cm}
\includegraphics[width=\columnwidth]{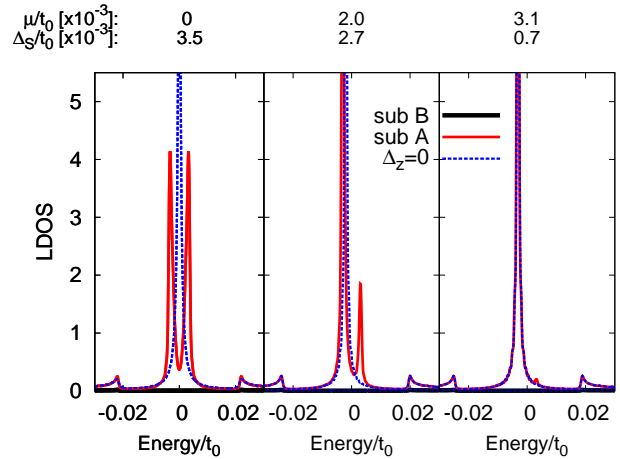}
%\vspace{-0.2cm}
\caption{ \label{ldop} (Color online) LDOS at the surface, on the A sublattice, for $U=-2.0t_0$ and three different dopings
considered before in Fig.~\ref{dopef}. The
normal state represented by the dashed line has been included for the corresponding values of doping $\mu$. $\Delta_S$ corresponds
to max$\{ \Delta_z \}$.
}
\end{figure}
In order to see the effect of external factors we  have considered  homogeneous doping, as well as an
inhomogeneous field-induced charge distribution,  along the z-direction.
The first case corresponds to graphite-intercalated compounds where dopant atoms are placed between the graphene
layers
while the inhomogeneous case can be easily realized in an experimental set-up where top and bottom electrodes  have opposite gate
voltages \cite{avetisyan_electric-field_2009,avetisyan_stacking_2010}.
Fig.~\ref{dopef} shows the evolution of the order parameter profile as a function of the homogeneous doping $\mu_z=\mu$ for a
fixed value of the pairing potential $U=-2.0t_0$.
Notice that, surface superconductivity becomes strongly suppressed as the doping shifts the Dirac point away from the Fermi energy. Total suppression occurs for doping lower than the  value of the order parameter at the surface.
Looking at the LDOS we found that this critical doping coincides with the extinction of the
coherence peaks and the rising of the single peak away form the Fermi energy.
Despite the fact that we do not find any relevant difference between homogeneous doping and surface only  doping in our self-consistent calculation, the  value of the critical doping is slightly higher than the one reported by the analytical results \cite{kopnin_high-temperature_2011} where the critical doping was found to be $\mu_{crit}$=$(2/3)\Delta_S$.

\begin{figure}[ttt]
%\vspace{-0.2cm}
\includegraphics[width=\columnwidth]{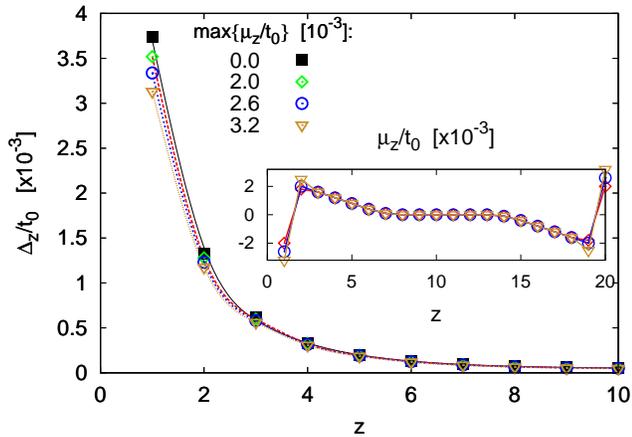}
%\vspace{-0.2cm}
\caption{\label{dopel} (Color online) Effect of inhomogeneous field-induced doping along the z-direction on the order parameter in a rhombohedral
multilayer graphene. Some cases previously shown in Fig.~\ref{dopef} are included for comparison. Inset: Doping profile along  the z-direction perpendicular to the graphene sheets.}
\end{figure}

\begin{figure}[ttt]
%\vspace{-0.2cm}
\includegraphics[width=\columnwidth]{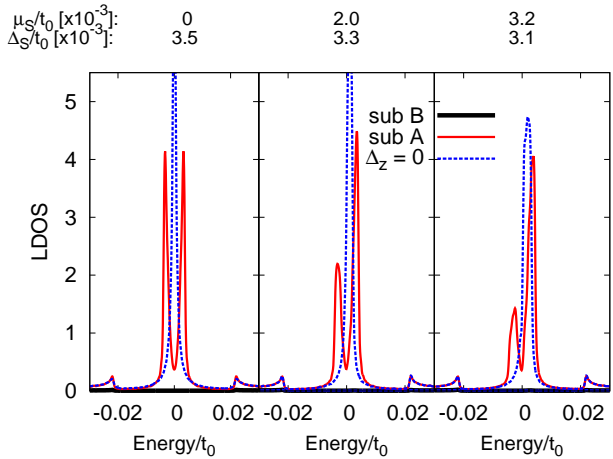}
%\vspace{-0.2cm}
\caption{\label{ldoel} (Color online) LDOS showing the effect of  field-induced doping along the z-direction (see inset in Fig.
\ref{dopel}) on  surface superconductivity, obtained for $U=-2.0t_0$ in ABC multilayer graphene. Dashed curves represent the
corresponding normal state LDOS.
}
\end{figure}

\begin{figure}[ttt]
%\vspace{-0.2cm}
\includegraphics[width=\columnwidth]{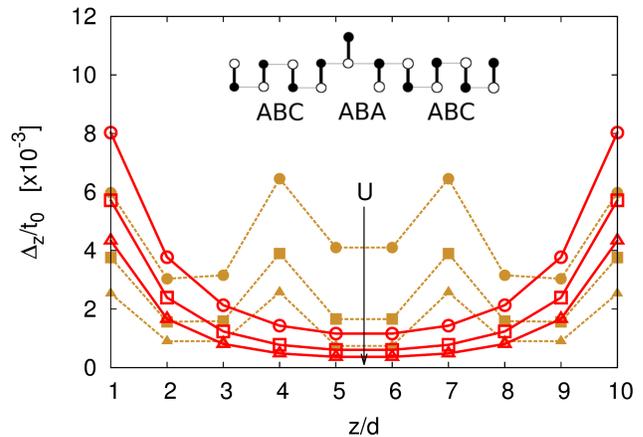}
%\vspace{-0.2cm}
\caption{\label{hyb} (Color online) Self-consistent order parameter profile for a multilayer graphene considering the stacking
configuration shown in the top part. Different point types correspond to different values of the pairing potential, $U/t_0=$ 2.16, 2.12 and 2.08,
where $U$ decrease in the direction pointed by the arrow. Open points represent the corresponding cases for an ordered ABC stacked multilayer graphene.
 }
\end{figure}

In addition to the homogeneous case we have also considered inhomogeneous doping, achieved when an electric field is applied perpendicular to the graphene sheets.
Following Ref.~\onlinecite{koshino_interlayer_2010} where the potential distribution due to an electric field was self-consistently calculated by taking into account screening effects,  we consider the z-dependent doping as described in the inset of Fig.~\ref{dopel}.
For comparison purposes we have considered the same pairing potential $U=-2.0t_0$, as we did in Fig.~\ref{dopef}, and three
different cases for which the doping at the surfaces is strongly suppressing the superconducting correlations in the homogeneous case.
Surprisingly, in contrast to  homogeneous doping, we found that the field-induced  doping suppresses only slightly the pair correlation at the
surface.
According to this result, the electric-field induced gap in
the inhomogeneous case is much lower than the value of the doping at the surface \cite{koshino_interlayer_2010}. Therefore, even
considering the same surface doping in both cases, superconducting correlations are weakly affected by an inhomogeneous  doping
configuration. Fig.~\ref{ldoel} shows how coherence peaks still survive for this doping configuration even if the value of the
doping at the surface is on the order of the surface order parameter (as obtained for zero-doping). This is in contrast with the
effect observed in Fig.~\ref{ldop} where for a similar level of doping causes a strong suppression of the superconducting
gap.\\
Finally, we consider multilayer graphene with hybrid stacking.
In order to shift the high-temperature surface superconductivity, observed in ABC, to the bulk of the structure, we
propose intercalated hybrid stacked graphite with few layers.
Numerical calculations considering stacking faults in multilayer graphene were recently reported \cite{kopnin_surface_2012},
showing that surface superconductivity survives at the interface between ABC and ABA stacking. However, superconductivity in the
bulk is still seen being suppressed when the thickness of the hybrid layers is large. By considering a few-layer structure where the external layers
have the ABC stacking configuration while the inner ones have ABA stacking, we expected a slight suppression of  surface superconductivity and an enhancement
of its bulk value. Fig.~\ref{hyb} shows self-consistent results obtained for the hybrid structure depicted in the top part of the figure. While
surface superconductivity appears suppressed as compared to the non-hybrid ABC multilayer, bulk correlation are also enhanced.
On the basis of these results we suggest that this kind of hybrid stacked multilayer structure, or more complex combinations, could support high temperature
superconductivity due to the interplay between surface superconductivity present in ABC stacking and the bulk superconductivity
preserved in the ABA case.
\section{Conclusions}
\label{conclusions}
By using a highly efficient GPU-based numerical procedure we solved self-consistently for the s-wave order parameter within a
mean-field approach for a tight-binding model of  ABA and ABC-stacked multilayer graphene. Main findings show that
a surface superconducting state appear when the multilayer is in the ABC stacking configuration. Opposite behavior is seen for the ABA stacking 
where bulk superconductivity is predominant but unstable below a certain critical pairing potential. 
The LDOS for surface sites shows peculiar 
coherence peaks and sublattice polarization, i.e. large LDOS in one sublattice and zero in the other. We showed that under homogeneous doping this
state is quickly suppressed. We extracted a critical doping which is slightly higher than the one reported previously based on an approximate analytical
study\cite{kopnin_high-temperature_2011}. In addition, we considered a field-induced 	inhomogeneous doping and showed that the superconducting correlations survive in this case for higher values of $\mu_S$.
Finally we pointed out the importance of hybrid stacking structures where surface superconductivity related to ABC stacking could be preserved
even in the bulk of the structure, suggesting a possible path for the survival of high temperature superconductivity.
\section*{Acknowledgements}
This work was supported by the Flemish Science Foundation (FWO-Vl) and the Methusalem funding of the Flemish Government.

\def\urlprefix{}
\def\url#1{}

%\bibliography{draft}% Produces the bibliography via BibTeX.


\end{document}